\documentclass[Journal]{IEEEtran}
\IEEEoverridecommandlockouts
\usepackage{amsmath,amssymb,amsfonts}
\usepackage{algorithmic}
\usepackage{graphicx}
\usepackage{xcolor}
\usepackage{soul}
\usepackage{multirow}
\usepackage{multicol}
\usepackage{url}
\usepackage{booktabs, tabularx, colortbl}
\usepackage{subfigure}
\usepackage{placeins}
\usepackage{bookmark}
\usepackage{hyperref}
\def\BibTeX{{\rm B\kern-.05em{\sc i\kern-.025em b}\kern-.08em
    T\kern-.1667em\lower.7ex\hbox{E}\kern-.125emX}}
\begin{document}

\title{Optimal Scheduling of Battery Storage Systems in the Swedish Multi-FCR Market Incorporating Battery Degradation and Technical Requirements
\thanks{The work presented in this paper is financially supported by the following projects: (i) ECom4Future FIWARE Driven Energy Communities for the Future - received funding from the Clean Energy Transition Partnership under Grant Agreement No. CETP22FP-2023-00316, and (ii) Implementation of the Vehicle to Grid Services in Sweden (PAVE) - received funding from Vinnova (Sweden's innovation agency) within the FFI program under the Authority's Dnr 2023-00785. 
\\The authors are with the Department of Electrical Engineering, Chalmers University of Technology, 412 96 Gothenburg, Sweden. (e-mail: nima.mirzaei@chalmers.se; khezrir@chalmers.se; mazadi@chalmers.se; david.steen@chalmers.se; tuan.le@chalmers.se}
}

 \author{\IEEEauthorblockN{Nima Mirzaei Alavijeh, Rahmat Khezri, \textit{Senior Member, IEEE,} Mohammadreza Mazidi, David Steen, and Le Anh Tuan, \textit{Member, IEEE}}}

\IEEEoverridecommandlockouts

\maketitle

\begin{abstract}
This paper develops a novel mixed-integer linear programming (MILP) model for optimal participation of battery energy storage systems (BESSs) in the Swedish frequency containment reserve (FCR) markets. The developed model aims to maximize the battery owner's potential profit by considering battery degradation and participation in multiple FCR markets, i.e., FCR in normal operation (FCR-N), and FCR in disturbances (FCR-D) for up- and down-regulations. Accordingly, a precise formulation of a detailed battery degradation model and adherence to the technical requirements of the Swedish FCR markets are incorporated into the developed model. To achieve more practical results, simulations are conducted based on one minute time step realistic data for the whole year 2022. The results show a potential profit of k€ 708 for a 1MW/1MWh BESS by participating in multi-FCR market. Analyzing the impact of considering degradation in the optimization problem has shown that the annual battery aging cost could decrease by 5\%-29\% without a significant effect on profit. The proposed model can be practically used by flexibility asset owners to achieve profitable and sustainable operation strategies that reduce battery degradation.
\end{abstract}


\IEEEpeerreviewmaketitle
\section*{Nomenclature}
\addcontentsline{toc}{section}{Nomenclature}
\subsection{Acronyms}
\begin{IEEEdescription}[\IEEEusemathlabelsep\IEEEsetlabelwidth{$V_1,V_2,V_3$}]
\item[BESS] Battery energy storage systems
\item[DA] Day-ahead
\item[EV] Electric vehicle
\item[FCR] Frequency containment reserve
\item[FCR-N] Frequency containment reserve in normal operation
\item[FCR-D] Frequency containment reserve in disturbances
\item[LER] Limited energy reservoirs 
\item[MILP] Mixed-integer linear programming 
\item[OM] Operation and maintenance 
\item[SoE] State of energy
\item[SoC] State of charge
\end{IEEEdescription}

\subsection{Indices and Sets}
\begin{IEEEdescription}[\IEEEusemathlabelsep\IEEEsetlabelwidth{$V_1,V_2,V_3$}]
\item[$d\in\mathcal{D}$] Index and set of days in a year
\item[$t\in\mathcal{T}_d$] Index and set of time steps in a day 
\item[$h\in \mathcal{H}_d$] Index and set of hours in a day 
\item[$\Xi$] Set of decision variables 
\end{IEEEdescription}
\subsection{Parameters and Functions}
\begin{IEEEdescription}[\IEEEusemathlabelsep\IEEEsetlabelwidth{$\rho^{\mathrm{N}}_h$, $\rho^{\mathrm{DD}}_h$,$\rho^{\mathrm{DU}}_h$}]
\item[$\mathcal{A}h_{t,d}$] Ah throughput of BESS at time step $t$ of day $d$ [Ah] 
\item[$\mathcal{B}^{\mathrm{BAT}}$] Net present value of battery [€]
\item[$\mathcal{B}_d^{\mathrm{DEG}}$] Cost of battery degradation on day $d$ [€]
\item[$\mathcal{B}_{t,d}^{\mathrm{CAL}}$, $\mathcal{B}_{t,d}^{\mathrm{CYC}}$] Calendar and cycle aging cost of battery degradation on time step $t$ of day $d$ [€]
\item[$C_d^{\mathrm{DA}}$] Cost of BESS in day-ahead spot market on day $d$ [€] 
\item[$C^{\mathrm{REP}}$] Cost of replacement of battery cells [€]
\item[$C^{\mathrm{REP}}$] Cost of operation and maintenance of BESS [€]
\item[$e_{t/h}^\mathrm{DR,X}$, $e_{t/h}^\mathrm{UR,X}$] Energy content: Per unit activated energy for down/up regulation for FCR-X service during $t$ or $h$ [h]
\item[$EOL$] The percentage of retained capacity at end of life of BESS [\%]
\item[$F_d$] Total profit of day $d$ [€]
\item[$f_{t,d}$] Grid frequency at time step $t$ of day $d$ [Hz]
\item[$f_n$] Nominal grid frequency [Hz]
\item[$f_{\mathrm{min}}^{\mathrm{N/D}}$] Minimum grid frequency for FCR-N or FCR-D activation [Hz]
\item[$f_{\mathrm{max}}^{\mathrm{N/D}}$] Grid frequency for full activation of FCR-N or FCR-D services [Hz]
\item[$\mathcal{I}_{t,d}$] C-rate of BESS at time step $t$ of day $d$
\item[$i$] Interest rate [\%] 
\item[$L$] Life time of BESS [Years]
\item[$\mathcal{K}_t$] Temperature at time step $t$ [K] 
\item[$\underline{P}$, $\overline{P}$] Minimum and maximum charger power (MW)
\item[$\underline{p}^{\Theta,X}$] Minimum bid size for FCR-X market, X $\in \{\mathrm{N, DU, DD}\}$ [MW]
\item[$r^{sv}$] Ratio of salvage cost to replacement cost 
\item[$R_d^{\mathrm{DA}}$] Revenue of BESS in day-ahead spot market on day $d$ [€] 
\item[$R_d^{\mathrm{FCR}}$] Revenue from FCR market participation on day $d$ [€] 
\item[$R_{h,d}^{\mathrm{X}}$] Revenue from FCR-X market on hour $h$ of day $d$, X $\in \{\mathrm{N, DU, DD}\}$ [€] 
\item[$\underline{S}$, $\overline{S}$] Minimum and maximum state-of-energy 
\item[$\mathcal{S}_{t,d}$] State of energy of BESS in time step $t$ of day $d$ [MWh] 
\item[$\mathcal{U}_{t,d}^{\mathrm{CAL}}$, $\mathcal{U}_{t,d}^{\mathrm{CYC}}$] Calendar and cycle aging of BESS on time step $t$ of day $d$ [\%]
\item[$\eta^{\mathrm{CH}}, \eta^{\mathrm{DS}}$] Charging, discharging efficiency
\item[$\rho^{\mathrm{spot}}_h$] Day-ahead spot market price at hour $h$ [€/MWh]
\item[$\rho^{\mathrm{tax}}$] Fixed electricity tax [€/MWh]
\item[$\rho^{\mathrm{g}}$] Grid utilization tariff [€/MWh]
\item[$\rho^{\mathrm{X}}_h$] Capacity reimbursement price for FCR-X market at hour $h$, X $\in \{\mathrm{N, DU, DD}\}$ [€/MW]
\item[$\rho_h^{\mathrm{UR}}$, $\rho_h^{\mathrm{DR}}$] Up/down regulation prices at hour $h$ [€/MWh]
\item[$\Delta p_{t,d}$] Deviation from baseline power due to the activation FCR services [MW]
\item[$\Delta t$] The length of the simulation time step [s]

\end{IEEEdescription}

\subsection{Variables}
\begin{IEEEdescription}[\IEEEusemathlabelsep\IEEEsetlabelwidth{$\rho^{\mathrm{N}}_h$, $\rho^{\mathrm{DD}}_h$,$\rho^{\mathrm{DU}}_h$}]
\item[$b_{t,d}^{\mathrm{CH/DS}}$] Binary variable for charge/discharge power of BESS in time step $t$ of day $d$
\item[$b_{h,d}^{\mathrm{CH/DS},\mathrm{bl}}$] Binary variable for baseline charge/discharge power of BESS at hour $h$ of day $d$
\item[$b_{h,d}^{X}$] Binary variable for participation in FCR-X market at hour $h$ of day $d$, X $\in \{\mathrm{N, DU, DD}\}$ 
\item[$p_{h,d}^{\Theta,\mathrm{X}}$] Power bid on FCR-X market on hour $h$ of day $d$, X $\in \{\mathrm{N, DU, DD}\}$ [MW] 
\item[$p_{t,d}^{\mathrm{CH/DS}}$] Charge/discharge power of BESS in time step $t$ of day $d$ [MW] 
\item[$p_{h,d}^{\mathrm{CH/DS}, \mathrm{bl}}$] Baseline (reference) charge/discharge power of BESS at hour $h$ of day $d$ [MW] 
\item[$p_{t,d}^{\mathrm{ND}}$, $p_{t,d}^{\mathrm{NU}}$] Activated up or down regulation for FCR-N service by BESS in time step $t$ of day $d$ [MW] 
\item[$p_{t,d}^{\mathrm{DD}}$, $p_{t,d}^{\mathrm{DD}}$] Activated up or down regulation for FCR-D services by BESS in time step $t$ of day $d$ [MW]

\end{IEEEdescription}

\section{Introduction}

\subsection{Motivation and Aim}
Frequency containment reserve (FCR) markets are one of the market-based tools used by the Nordic electricity system operators for regulating grid frequency. FCR services comprises three products: FCR in normal operation (FCR-N) and FCR in disturbances (FCR-D) for up and down regulations, to manage small and larger frequency fluctuations, respectively \cite{modig2022overview}.

A recent market trend is the increasing participation of distributed flexibility resources. Flexibility asset owners involved in providing FCR must maintain fast response times and engage in shorter, but more frequent activation events. Accordingly, battery energy storage systems (BESSs) are considered ideal FCR providers because, as inverter-coupled resources, they can deliver fast power response despite their limited energy capacity \cite{8864014}. It was shown by Hu et al. \cite{hu2022potential} that BESSs generate the highest revenue by providing various FCRs in most European electricity markets. However, to ensure a profitable scheduling strategy for BESSs, it is crucial to consider the market requirements and precisely model the costs associated with battery degradation. 

This paper aims to develop an optimal scheduling model for the efficient participation of BESSs in the Swedish energy and FCR markets. The objective is to maximize profit while considering battery degradation accurately and meeting all technical requirements of the Swedish markets. 

\subsection{Literature Review}

By analyzing the literature on the application of BESSs in the frequency regulation of power systems, two distinct modeling approaches can be identified, i.e., control-based and market-based approaches. The first approach focuses on the optimal control of BESSs to effectively regulate frequency. In \cite{8999747}, a distributed control strategy was proposed for multiple BESSs to regulate frequency in the power system with high penetration of renewable generation. In \cite{10045057}, an online frequency regulation strategy based on the Lyapunov optimization technique was proposed. In \cite{9445583}, it was proposed that the frequency regulation signal can be divided into a slow component, directed to synchronous generators, and a fast component, directed to BESSs. Authors of \cite{9102420} proposed a robust control strategy to manage distributed BESSs for frequency regulation. While these papers focused on developing control strategies for BESSs to regulate frequency, they did not address the optimization of profits that can be obtained by participating in the ancillary service market. 

\begin{figure*}[ht]
    \centering
    \includegraphics[width=0.95\linewidth]{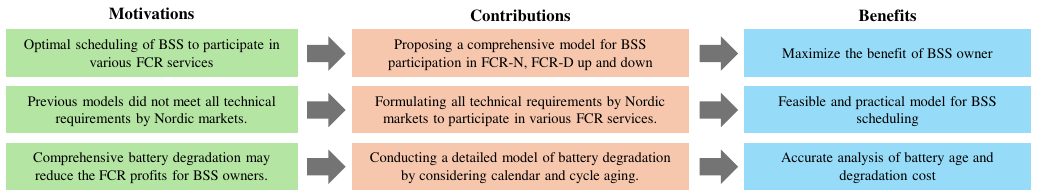}
    \caption{Motivations, contributions, and benefits of the conducted study.}
    \label{fig:contributions}
\end{figure*}

The participation of BESSs in ancillary service markets is investigated in the second approach. An example of such an approach is \cite{9103134}, where a bidding strategy and online control methodology for BESSs was proposed to enable participation in both the day-ahead energy market and the FCR-N market. While the battery degradation was modeled, participation in the FCR-D up/down market was not considered. Furthermore, the proposed model was nonlinear which adds computational burden and makes finding the global optimum solution challenging. In \cite{9832632}, a two-stage stochastic optimization model was developed for the optimal bidding of BESSs in both energy and ancillary service markets. While the model effectively addressed participation in the multi-FCR market, it fell short in full accounting for market requirements. Additionally, the accuracy of battery degradation modeling in the proposed framework was deemed simplistic. In \cite{9508854}, electric vehicles (EVs) were aggregated as a BESS, and a two-level optimization model was proposed to determine their optimal participation in energy and ancillary service markets. In the proposed model, sole participation in the FCR-N market was considered, while factors such as battery degradation and market requirements were neglected. In \cite{9737408}, robust and stochastic methods were used to model the participation of aggregated EVs in both the energy and ancillary service markets. In the developed model, a constant annual activation ratio was calculated for FCR using real historical data and incorporated into the supply-demand balance equation as a random variable. While their developed model considered multi-FCR market participation, it was not developed with the technical requirements of Nordic FCR markets, and the battery degradation is not a detailed model. The authors of \cite{subroto2022bess} have introduced an optimal model for sizing and scheduling BESSs to maximize income from participating in the energy and FCR-N markets. However, the proposed model employed a simplified degradation model for the battery and overlooked the FCR market requirements.

To ensure a profitable strategy for BESSs, the battery degradation should be precisely modeled. This is because, during FCR-N activation, BESS is used as a power buffer, alternating between charging and discharging cycles. This can potentially result in substantial energy exchanges between the BESS and the grid, thereby accelerating battery degradation in addition to high conversion losses \cite{thingvad2022economic}. The impact of providing FCRs on the degradation of BESSs has been examined in \cite{jacque2022influence}. The results indicate that BESSs have been subjected to many cycles with a low average cycle depth providing FCRs, leading to a dominance of calendar aging over cyclic aging. In \cite{krupp2023operating}, degradation has been identified as the most critical factor affecting the profitability of BESSs. Although energy exchange between the BESS and the grid is significantly lower in FCR-D up/down markets compared to the FCR-N market, FCR-N's bid price is higher than the FCR-D up/down in most of the time. Therefore, to maximize the profitability of BESSs, it is essential to consider participation in multi-FCR markets while accounting for degradation. As demonstrated in \cite{thingvad2022economic}, selecting the best combination of FCRs in each hour of BESS scheduling can increase the profitability by 22\% compared to delivering pure FCR-N. However, modeling the participation of BESSs in multi-FCR markets is a burdensome and demanding task nowadays due to the complicated and essential requirements of FCR markets. As investigated in \cite{koltermann2023power}, neglecting market requirements may lead to situations where balancing the BESS is insufficient to provide the FCRs, resulting in penalties.

After a detailed review of the literature and thorough focus on relevant research works, owing to the best of our knowledge, it has been found that no work has been dedicated to providing an optimal scheduling model for BESSs in the multi-FCR market considering battery degradation and meeting all technical requirements of the Swedish markets. 

\subsection{Approach and Contributions}

To fill the research gaps, this study develops a new optimal BESS scheduling model for multi-market FCR participation by considering all technical requirements of the Swedish market and comprehensive battery degradation. The proposed model is formulated with respect to a potential stacked hourly participation in FCR markets while optimizing the purchased power from the DA spot market acting as an optimal reference power for delivering FCR services. From a practical point of view, the developed model in this study can be utilized by flexibility asset owners in obtaining more sustainable and yet profitable operation strategies for BESSs. 
The main contributions of this study can be summarized as follows:
\begin{itemize}
    \item Formulating a MILP model for optimal scheduling of BESSs in DA spot and multi-FCRs markets which can be solved efficiently through optimization software.
    \item Incorporating all technical requirements of the Swedish FCR market in the developed model and solving the optimal scheduling problem based on one minute realistic data for one year.
    \item Applying a detailed model of battery degradation in the developed model to ensure the profitability of the model.
\end{itemize}

The proposed model enables time-efficient analyses on how the maximum potential profit and participation strategy of BESS are impacted by input parameters and model settings such as the inclusion of battery degradation, and the inclusion of FCR markets technical requirements. By proposing this model, the authors are pursuing the following objectives: 
\begin{itemize}
    \item Modeling charge and discharge powers as variables linked to the bid size and the activation criteria, enabling degradation to be considered in the optimization.
    \item Including endurance and power requirements from the latest technical requirements for the participation of limited energy reservoirs in FCR markets.
    \item Enabling the possibility of stacked participation in multiple FCR markets at each hour.
    \item Optimizing the reference power considering both the DA spot and FCR markets.
\end{itemize}

The motivations, contributions, and benefits of the developed model in this study are presented in Figure \ref{fig:contributions}.

\subsection{Paper Organization}
The rest of the paper is laid out as follows. The frequency containment reserve and market requirements are presented in Section \ref{sec:Swedish_fcr}. The model formulation considering the battery degradation is described in Section \ref{sec:formulation}. Case study and input data are elaborated in Section \ref{sec:casestudy_data}. The results and discussion are given in Section \ref{sec:results_discussion} and conclusions are provided in Section \ref{sec:conclusions}.

\section{Swedish Frequency Containment Reserves}\label{sec:Swedish_fcr}
In Sweden, there are three services for FCR market: FCR-N, FCR-D up, and FCR-D down. There are differences between regulations and technical requirements for these FCR services. The main aspects of various FCR services are summarized in Table \ref{tab:FCR_requirements}. For a realistic model formulation, applying these aspects is essential. 

\begin{table}[htbp]
\centering
\caption{Comparison of market and technical requirements of FCR services \cite{entsoe_tech_req,InformationDifferentAncillary2024}}
\label{tab:FCR_requirements}
\newcolumntype{Y}{>{\centering\arraybackslash}X}
\begin{tabularx}{\linewidth}{@{}p{3.65cm}XXX@{}}
\toprule \toprule
                                          & FCR-N  & FCR-D up  & FCR-D down   \\ \midrule
                                         \centering \textbf{Market}&&&\\ \cmidrule{1-1}
                                         Trading time  & Day-ahead  & Day-ahead & Day-ahead \\
                                         Remuneration  & Capacity and energy & Capacity  & Capacity \\
                                         Min bid size & 0.1 MW & 0.1 MW  & 0.1 MW \\ \cmidrule{1-1}
\centering\textbf{Technical requirements}&&& \\ \cmidrule{1-1}
 Required power upwards & \textrm{$1.34\cdot p_{h}^{\mathrm{\Theta,N}}$} & $p_{h}^{\mathrm{\Theta,DU}}$ & $0.2 \cdot p_{h}^{\mathrm{\Theta,DD}}$ \\
                                         Required power downwards  & \textrm{$1.34\cdot p_{h}^{\mathrm{\Theta,N}}$} & $0.2 \cdot p_{h}^{\mathrm{\Theta,DU}}$ & $p_{h}^{\mathrm{\Theta,DD}}$                         \\
                                         Required endurance upwards   &  1h & \textrm{$\frac{1}{3}$}h & 0 \\
                                         Required endurance downwards &  1h &  0 &  \textrm{$\frac{1}{3}$}h \\
                                         Activation & Automatic (Figure \ref{fig:droop}) & Automatic (Figure \ref{fig:droop}) & Automatic (Figure \ref{fig:droop}) \\ \bottomrule \bottomrule
\end{tabularx}
\end{table}

From a market perspective, the difference between the FCR services concerns their remuneration. All three services are remunerated for their cleared bid capacity based on pay-as-cleared prices \cite{InformationDifferentAncillary2024}. In addition, FCR-N is remunerated also based on the activated energy. This is because of the relatively higher energy throughput in FCR-N services. Up-regulation energy is compensated with up-regulation prices, while the service provider is charged with down-regulation prices for the down-regulation energy. 

To calculate realistic bidding and profit, technical requirements from the FCR markets need to be considered. ENTSO-e is the European association for the cooperation of transmission system operators that has provided the technical requirements \cite{entsoe_tech_req} for limited energy reservoirs (LER) such as EVs and batteries. The power requirements are formulated based on the accepted bid $P_h^\mathrm{\Theta,X}$. The endurance requirements indicate how long the service provider must be able to provide the accepted bid at a full activation scenario. 

\begin{figure}[htbp]
     \centering
     \includegraphics[width=\linewidth]{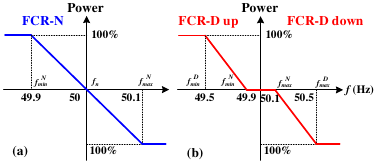}
     \caption{FCR activation power based on grid frequency: (a) FCR-N, (b) FCR-D}
     \label{fig:droop}
 \end{figure}

The activation of FCR services is based on the specified droop curves shown in Figure \ref{fig:droop}. These curves should be strictly followed for the accepted bid size. There are exceptions for LERs in case their state of charge (SOC) goes below or above certain limits specified in \cite{entsoe_tech_req}. In case these limits are passed, the resource should deviate from the droop curves to bring back the SOC to the acceptable ranges. In such cases, the service provider will not be remunerated since it could not provide the service. These exceptions are not considered in this study and instead a safety margin is implemented for the SOC by its lower and upper bounds.

Considering the market and technical requirements in the model is essential for a realistic estimation of bidding behavior and profitability of services. In Section \ref{sec:formulation}, a novel MILP formulation is proposed for implementing these requirements.

\section{Model Formulation}\label{sec:formulation}
The developed optimization model in this study is presented in this section. The model formulation is presented in different subsections representing the objective function, and constraints related to the battery’s physical limits, FCR participation rules and requirements, and battery degradation model. 

\subsection{Objective Function}
The objective function of the BESS scheduling problem is to maximize the daily profit given by:  
\begin{subequations}
\label{eq:OF}
\begin{align}
f & = \max_{\Xi} \quad F_d \\
& = \max_{\Xi} \quad (R_d^{\mathrm{DA}} + R_d^{\mathrm{FCR}} - C_d^{\mathrm{DA}} - C_d^{\mathrm{DEG}}), \quad \forall d \in \mathcal{D}, \notag\\
& \Xi =\{p_{h,d}^{\mathrm{CH,bl}}, p_{h,d}^{\mathrm{DS,bl}}, p_{t,d}^{\mathrm{CH}}, p_{t,d}^{\mathrm{DS}},\\
&\quad \quad \quad p_{h,d}^{\mathrm{\Theta,N}}, p_{h,d}^{\mathrm{\Theta,DU}}, p_{h,d}^{\mathrm{\Theta,DD}}  \mid h \in \mathcal{H}_d, t \in \mathcal{T}_{d}\} \notag, \quad  \forall d \in \mathcal{D} 
\end{align}
\end{subequations}
where $\mathcal{D}$ is the set of all the days in a year, $\mathcal{T}_d$ is the set of all the timestamps in day d with a length of lower than one hour, $\mathcal{H}_d$ is the set of all hours in day $d$. $R_d^{\mathrm{DA}}$ is the revenue from DA spot market through discharging, $R_d^{\mathrm{FCR}}$ is the revenue from FCR participation, $C_d^{\mathrm{DA}}$ is the cost in DA spot market through charging, and $C_d^{\mathrm{DEG}}$ is the cost for battery degradation. The decision variables of the optimization problem are the hourly purchased and sold power in the DA spot market ($p_{h,d}^{\mathrm{CH,bl}}, p_{h,d}^{\mathrm{DS,bl}}$) that are acting as the reference point (baseline) for the FCR services, charging and discharging power of the battery ($p_{t,d}^{\mathrm{CH}}$, $p_{t,d}^{\mathrm{DS}}$), power bid to participate in FCR-N, FCR-D up and FCR-D down ($p_{h,d}^{\Theta,\mathrm{N}}$, $p_{h,d}^{\Theta,\mathrm{DU}}$, $p_{h,d}^{\Theta,\mathrm{DD}}$). To avoid repetition, the indices related to the day are not written in the rest of the equations.

The DA spot market revenue is formulated based on hourly spot market price ($\rho_h^{\mathrm{spot}}$), and discharging power ($p_{h}^{\mathrm{DS,bl}}$). 
\begin{equation}
\label{eq:rev_DA}
R^{\mathrm{DA}} = \sum_{h \in \mathcal{H}}{p_{h}^{\mathrm{DS,bl}} \cdot (\rho_h^{\mathrm{spot}}+\rho^{\mathrm{tax}})}
\end{equation}

The revenue for multi-market FCR participation contains three parts of revenues for FCR-N, down-regulation FCR-D, and up-regulation FCR-D. 
\begin{equation}
\label{eq:rev_FCR}
R^{\mathrm{FCR}} = \sum_{h \in \mathcal{H}}{R_h^\mathrm{N} + R_h^{\mathrm{DD}} + R_h^{\mathrm{DU}}}
\end{equation}
The FCR-N revenue includes three terms. The first term is the capacity reimbursement for power bid. The capacity is reimbursed according to the FCR-N market price ($\rho_h^{\mathrm{N}}$) and the amount of power bid ($p_{h}^{\Theta,\mathrm{N}}$). The second and third terms are the energy compensation for upregulation and downregulation during the FCR-N participation. These terms are calculated based on the concept of energy content introduced by \cite{thingvad2022economic}. Energy content represents the per unit activated energy for up or down ward regulations within a specific timestep $t$ or a specific hour $h$. It is pre-calculated based on the frequency and droop settings. Multiplying energy content ($e_{h}^{\mathrm{UR/DR,N}}$), FCR-N power bid, and up and down regulation prices ($\rho_h^{\mathrm{UR/DR}}$) at each hour will provides the hourly energy compensation for FCR-N services
\begin{equation}
\label{eq:rev_FCR-N}
R_h^\mathrm{N} = p_h^{\Theta,\mathrm{N}} \cdot \left(\rho_h^\mathrm{N}\ +\rho_h^{\mathrm{UR}} \cdot e_h^{\mathrm{UR,N}} - \rho_h^{\mathrm{DR}}\cdot e_h^{\mathrm{DR,N}}\right),\quad \forall h 
\end{equation}

The revenues for FCR-D down and up are only based on capacity reimbursement according to FCR-D up and down prices and the power bid.
\begin{equation}
\label{eq:rev_FCR-D down}
R_h^\mathrm{DD}=\rho_h^\mathrm{DD}\ p_h^{\Theta,\mathrm{DD}},\quad \forall\ h	
\end{equation}
\begin{equation}
\label{eq:rev_FCR-D up}
R_h^\mathrm{DU}=\rho_h^\mathrm{DU}\ p_h^{\Theta,\mathrm{DU}},\quad \forall\ h	
\end{equation}

The expenditure for charging in the DA spot market is formulated based on the hourly spot price, grid utilization cost ($\rho^g$), electricity tax ($\rho^{\mathrm{tax}}$), and charging power.
\begin{equation}
\label{eq:cost_DA}
C^{\mathrm{DA}}=\sum_{h\in\mathcal{H}}{p_h^{\mathrm{CH,bl}}\left(\rho_h^{\mathrm{spot}}+\rho^g+\rho^{\mathrm{tax}}\right)}	
\end{equation}

The cost of battery degradation is the sum of cycle aging cost ($\mathcal{B}_t^{\mathrm{CYC}}$) and calendar aging cost ($\mathcal{B}_t^{\mathrm{CAL}}$):
\begin{equation}
\label{eq:C_deg_generic}
C^{\mathrm{DEG}}=\sum_{t\in T}\left(\mathcal{B}_t^{\mathrm{CAL}}+\mathcal{B}_t^{\mathrm{CYC}}\right)\ 
\end{equation}

\subsection{Battery Constraints}

The battery model contains several constraints related to power and state of energy (SoE) in the optimization model. The charging and discharging power are either zero or between $\underline{P}$ and $\bar{P}$. In fact, the power cannot exceed the charger’s maximum power limit at any time. On the other hand, due to low efficiency in low power, the chargers do not allow charging or discharging with a power lower than the minimum limit.
\begin{equation}
\label{eq:bes_p_pbl_bounds}
p_h^\mathrm{CH,bl},\ p_h^\mathrm{DS,bl},P_t^\mathrm{CH},P_t^\mathrm{DS}\in\left\{0\right\}\cup\left[\underline{P} , \overline{P}\, \right],\quad \forall h
\end{equation}

To avoid simultaneous charging and discharge, and to implement (\ref{eq:bes_p_pbl_bounds}), binary variables ($b_t^\mathrm{CH}$, $b_t^\mathrm{DS}$, $b_h^\mathrm{CH,bl}$, $b_h^\mathrm{DS,bl}$) are considered. The sum of the first pair and the sum of the second of binary variables should be less than or equal to one at their respective timestep. 
\begin{equation}
\label{eq:bes_p_bounds_constraints}
b_t^\mathrm{CH/DS} \cdot \underline{P}\le p_t^\mathrm{CH/DS} \le b_t^\mathrm{CH/DS} \cdot \overline{P},\quad \forall t
\end{equation}
\begin{equation}
\label{eq:bes_pbl_bounds_constraints}
b_h^\mathrm{CH/DS,bl} \cdot \underline{P} \le p_h^\mathrm{CH/DS,bl} \le b_h^\mathrm{CH/DS,bl} \cdot \overline{P},\quad  \forall h
\end{equation}
\begin{equation}
\label{eq:bes_b_p_constraints}
b_t^\mathrm{CH} + b_t^\mathrm{DS}\le 1,\quad  \forall t
\end{equation}
\begin{equation}
\label{eq:bes_b_pbl_constraints}
b_h^\mathrm{CH,bl} + b_h^\mathrm{DS,bl}\le 1,\quad \forall h
\end{equation}

The SoE ($\mathcal{S}_{t}$) is formulated based on the SoE from the previous timestamp ($\mathcal{S}_{t-1,d}$), charging and discharging baseline power, and the up and down regulation energy corresponding to FCR services.
\begin{align}
\label{eq:bes_soe}
\mathcal{S}_t = \mathcal{S}_{t-1} &+ \left(P_h^\mathrm{CH,bl}\cdot\eta^\mathrm{CH} -P_h^\mathrm{DS,bl}/\eta^\mathrm{DS}\right) \cdot \Delta t \\
&+ P_{h_t}^\mathrm{\Theta,N} \cdot \left(e_t^\mathrm{DR, N}-e_t^\mathrm{UR,N}\right)\notag\\
&+ P_{h_t}^\mathrm{\Theta,DD}\cdot e_t^\mathrm{DR,DD}\notag\\
&- P_{h_t}^\mathrm{\Theta,DU}\cdot e_t^\mathrm{UR,DU},\quad \forall t \notag
\end{align}
\begin{equation}
\label{eq:bes_soe_bounds}
\underline{\mathcal{S}}\le\mathcal{S}_t\le\overline{\mathcal{S}},\quad \forall t
\end{equation}
where $\eta^\mathrm{CH}$ and $\eta^\mathrm{DS}$ are the charging and discharging efficiencies, and $e_t^\mathrm{UR/DR, X}$  is the energy content of up- or down-regulation activation for different FCR markets. Hour $h_t$ represents the corresponding hour to timestamp $t$.

\subsection{FCR participation constraints}
Several constraints need to be considered to satisfy the technical requirements and physical limitations related to stacked multi-FCR market participation.
The activated power for delivering FCR-N service is formulated based on the frequency deviations of the power system. It follows the droop control shown in Figure \ref{fig:droop}.
\begin{align}\label{eq:droop_ND}
p_t^\mathrm{ND}=
    \begin{cases}
        p_h^\mathrm{\Theta,N}, & f_t>f_{\max}^\mathrm{N}\\
        p_h^\mathrm{\Theta,N}\left(\frac{f_t - f_n}{f_{\max}^\mathrm{N} - f_n}\right), & f_n \le f_t \le f_{\max}^\mathrm{N}\\
        0, & f_t<f_n 
    \end{cases}
    ,\quad \forall t,
\end{align}
\begin{align}\label{eq:droop_NU}
p_t^\mathrm{NU} =
\begin{cases}
    0 ,& f_t>f_n\\
    p_h^\mathrm{\Theta,N} \left(\frac{f_t - f_n}{f_{\min}^\mathrm{N} - f_n}\right), & f_{\min}^\mathrm{N} \le f_t\le f_n\\
    p_h^\mathrm{\Theta,N}, & f_t< f_{\min}^\mathrm{N}
\end{cases}
,\quad \forall t,
\end{align}
where $f_t$ is the measured frequency of the system in Hz in timestamp $t$, $f_n$ is the nominal frequency (50 Hz), $f_{\max}^\mathrm{N}$ and $f_{\min}^\mathrm{N}$ are the maximum and minimum ranges of frequency for FCR-N participation, 50.1 Hz and 49.9 Hz, respectively.

The activated power for delivering FCR-D markets is also formulated based on the frequency deviations of the power system according to Figure \ref{fig:droop}.
\begin{align}\label{eq:droop_DD}
p_t^\mathrm{DD}=
\begin{cases}
p_h^\mathrm{\Theta,DD},& f_t > f_{\max}^\mathrm{D}\\
p_h^\mathrm{\Theta,DD}\left(\frac{f_t - f_{\max}^\mathrm{N}}{f_{\max}^\mathrm{D} - f_{\max}^\mathrm{N}}\right),& f_{max}^\mathrm{N}\le f_t\le f_{\max}^\mathrm{D}\\
0,& f_t < f_{\max}^\mathrm{N}
\end{cases}
,\quad \forall\ t,
\end{align}
\begin{align}\label{eq:droop_DU}
p_t^\mathrm{DU}=
\begin{cases}
0, & f_t > f_{\min}^\mathrm{N}\\
p_h^\mathrm{\Theta,DU}\left(\frac{f_t-f_{\min}^\mathrm{N}}{f_{\min}^\mathrm{D} - f_{\min}^\mathrm{N}}\right),& f_{\min}^\mathrm{D}\le f_t \le f_{\min}^\mathrm{N}\\
p_h^\mathrm{\Theta,DU}, & f_t<\ f_{\min}^\mathrm{D}
\end{cases}
,\quad \forall\ t,
\end{align}
where $f_{\max}^\mathrm{D}$ and $f_{\min}^\mathrm{D}$ are the maximum and minimum ranges of frequency for FCR-D participation, 50.5 Hz and 49.5 Hz, respectively.

The activated power for FCR services should be delivered with respect to the original plan of the battery. This is shown in (\ref{eq:p_and_delta_p}) where the original plan is the baseline power. The baseline is the traded energy in the DA spot market in the context of this model. The deviation from the baseline is shown by $\Delta p_t$. A positive $\Delta p_t$ corresponds to down-regulation and negative to up-regulation because load convention is used for the power.
\begin{align}
    \label{eq:p_and_delta_p}
    \Delta p_t &= p_t - p_{h_t}^\mathrm{bl}\\ &= \left(p_t^\mathrm{CH} - p_t^\mathrm{DS}\right)-\left(p_{h_t}^\mathrm{CH,bl} - p_{h_t}^\mathrm{DS,bl}\right), \quad \forall t \notag
\end{align}

When bidding in the FCR markets, $\Delta p_{t}$ should be strictly equal to the specified power levels from the droop controls:
\begin{equation}
    \label{eq:delta_p}
\Delta P_t=P_t^\mathrm{ND} + P_t^\mathrm{DD} - P_t^\mathrm{NU} - P_t^\mathrm{DU},\quad \forall t.
\end{equation}

The lower and upper bounds related to the bids need to be defined. Power bids for each market should not exceed double the maximum power of the battery. The upper bound is double the maximum power because the battery can, for example, go from full charge to full discharge power. The minimum power is also considered for each market participation ($\underline{p}^\mathrm{\Theta,N}$, $\underline{p}^\mathrm{\Theta,DU}$, and $\underline{p}^\mathrm{\Theta,DD}$). Binary variables ($b_t^\mathrm{N}$, $b_t^\mathrm{DD}$, and $b_t^\mathrm{DU}$) are added to implement the minimum power bid:
\begin{equation}
    \label{eq:up_low_bound_FCR-Nbid}
 b_h^\mathrm{N} \cdot \underline{p}^\mathrm{\Theta, N}\le p_h^\mathrm{\Theta, N}\le b_h^\mathrm{N}\cdot\overline{P}, \quad \forall h,	
\end{equation}
\begin{equation}
    \label{eq:up_low_bound_FCR-D downbid}
 b_h^\mathrm{DD} \cdot \underline{p}^\mathrm{\Theta, DD}\le p_h^\mathrm{\Theta,DD}\le b_h^\mathrm{DD}\cdot2\overline{P}, \quad \forall h,	
\end{equation}
\begin{equation}
\label{eq:up_low_bound_FCR-D upbid}
 b_h^\mathrm{DU} \cdot \underline{p}^\mathrm{\Theta, DU}\le p_h^\mathrm{\Theta,DU} \le b_h^\mathrm{DU}\cdot2\overline{P}, \quad \forall h.	
\end{equation}

\begin{figure}[htbp]
    \centering
    \includegraphics[width=\linewidth]{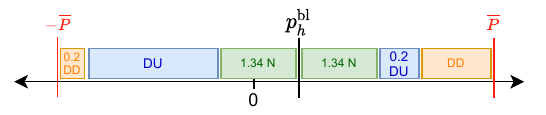}
    \caption{Limits on FCR capacity bids considering the baseline and ENTSOe requirements on power. N: $P_h^\mathrm{\Theta,N}$, DU: $P_h^\mathrm{\Theta,DU}$, DD: $P_h^\mathrm{\Theta,DD}$}
    \label{fig:power_req}
\end{figure}

To have results closer to real life, technical requirements for the FCR markets (Table \ref{tab:FCR_requirements}) need to be considered. 
The power requirements are transferred to the physical limitations of the battery in Figure \ref{fig:power_req} using the maximum charge and discharge capacity ($\overline{P}$) and the baseline ($p_h^\mathrm{bl}$) which is the reference point for providing the up- or down-regulation. The figure can be formulated as constraints provided in (\ref{eq:power_req_Up}) and (\ref{eq:power_req_Down}).
\begin{equation} \label{eq:power_req_Up}
1.34p_h^\mathrm{\Theta,N} + p_h^\mathrm{\Theta,DU} + 0.2p_h^\mathrm{\Theta,DD} \le \overline{P} + p_h^\mathrm{bl}, \quad \forall h
\end{equation}
\begin{equation}\label{eq:power_req_Down}
1.34p_h^\mathrm{\Theta,N} + p_h^\mathrm{\Theta,DD} + 0.2p_h^\mathrm{\Theta,DU} \le \overline{P} - p_h^\mathrm{bl}, \quad \forall h
\end{equation}

\begin{figure}
    \centering
    \includegraphics[width=\linewidth]{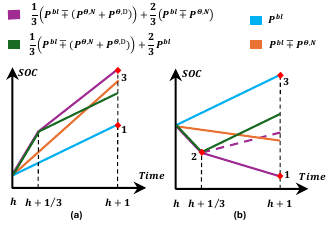}
    \caption{Examples of critical activation cases for endurance requirement when reference power is at charging. (a) downwards regulation, (b) upwards regulation}
    \label{fig:endurance_req}
\end{figure}

The energy requirements (i.e., endurance) can be implemented for the worst-case scenarios for SoC at each hour. The potential extreme scenarios are visualized in Figure \ref{fig:endurance_req} for up- and downwards activation when the reference power is at charging. A similar analysis can be done for a reference power at discharging. These most critical points are marked with red circles and can occur in the following scenarios:
 \begin{enumerate}
     \item If the service is not activated at all and the battery is locked to follow its baseline: equation (\ref{eq:endur_req_scn1})
     \item If both FCR-N and FCR-D services are activated for 20 minutes: equation (\ref{eq:endur_req_scn2})
     \item If FCR-N and FCR-D services are simultaneously activated for 20 minutes and FCR-N for the remaining 40 minutes within the hour: equation (\ref{eq:endur_req_scn3})
 \end{enumerate}
\begin{equation}\label{eq:endur_req_scn1}
\underline{\mathcal{S}} \le S_{h-\Delta t} + P_h^\mathrm{bl} \le \overline{\mathcal{S}}, \quad \forall h
\end{equation}
\begin{subequations}
\label{eq:endur_req_scn2}
\begin{align}
    \underline{\mathcal{S}} \le \mathcal{S}_{h-\Delta t} +\left(p_h^\mathrm{bl}+p_h^\mathrm{\Theta,N}+p_h^\mathrm{\Theta,DD}\right) \cdot \frac{1}{3} \le \overline{\mathcal{S}}, \quad \forall h\\
    \underline{\mathcal{S}} \le \mathcal{S}_{h-\Delta t} + \left(p_h^\mathrm{bl} - p_h^\mathrm{\Theta,N} - p_h^\mathrm{\Theta,DU}\right) \cdot\frac{1}{3} \le \overline{\mathcal{S}}, \quad \forall h
\end{align}
\end{subequations}
\begin{subequations}\label{eq:endur_req_scn3}
\begin{align}
\underline{\mathcal{S}} \le \mathcal{S}_{h-\Delta t} + p_h^\mathrm{bl} + p_h^\mathrm{\Theta,N} + \frac{1}{3} \cdot p_h^\mathrm{\Theta,DD}\le \overline{\mathcal{S}}, \quad \forall h	\\
\underline{\mathcal{S}} \le \mathcal{S}_{h-\Delta t} + p_h^\mathrm{bl} - p_h^\mathrm{\Theta,N} - \frac{1}{3} \cdot p_h^\mathrm{\Theta,DU}\le \overline{\mathcal{S}}, \quad \forall h
\end{align}
\end{subequations}

\subsection{Battery degradation}
The costs of cycle aging and calendar aging of the battery are calculated based on the degradation of the battery due to cycle aging ($\mathcal{U}_{t,d}^\mathrm{CYC}$) and calendar aging ($\mathcal{U}_{t,d}^\mathrm{CAL}$), percentage of retained capacity at end-of-life (EOL), and the net present value of the battery ($\mathcal{B}^\mathrm{BAT}$).
\begin{equation}
    \mathcal{B}_t^\mathrm{CAL}=\mathcal{B}^\mathrm{BAT}\frac{\mathcal{U}_t^\mathrm{CAL}(\%)}{100\%-\mathrm{EOL}(\%)},\quad \forall\ t,	
\end{equation}
\begin{equation}
    \mathcal{B}_t^\mathrm{CYC}=\mathcal{B}^\mathrm{BAT}\frac{\mathcal{U}_{t,d}^\mathrm{CYC}(\%)}{100\%-\mathrm{EOL}(\%)},\quad \forall\ t.
\end{equation}

Typically, the percentage of retained capacity at the EOL of the battery is considered as 80\% \cite{10488452}. The net present value of the battery is calculated based on the lifetime of the battery (L), interest rate (i), cost of replacement ($C^\mathrm{REP}$), cost of operation and maintenance ($C^\mathrm{OM}$), and the ratio of salvage cost to replacement cost ($r^\mathrm{sv}$).
\begin{equation}
\mathcal{B}^\mathrm{BAT}=\left(1-r^\mathrm{sv}\right)\ C^\mathrm{REP}\frac{1}{\left(1+i\right)^L}+C^\mathrm{OM}\frac{\left(1+i\right)^L-1}{{\alpha\left(1+i\right)}^L}
\end{equation}

The battery degradation model is taken from \cite{wang2014degradation} where an experimental model was developed for a Lithium-Nickel-Manganese-Cobalt + Lithium-Manganese oxide battery. The authors derived nonlinear empirical aging models for the same battery where the calendar aging depends on SOC, temperature ($\mathcal{K}_t$), and battery age in days ($\mathcal{Q}$), and the cycle aging depends on temperature, C-rate ($\mathcal{I}_t$), and Ah-throughput (${\mathcal{A}h}_t$). 
\begin{equation}
\mathcal{U}_t^\mathrm{CAL} = \mathcal{F}\left(\mathcal{S}_t,\mathcal{K}_t,\mathcal{Q}\right), \quad \forall t
\end{equation}
\begin{equation}
\mathcal{U}_t^\mathrm{CYC} = \mathcal{F}\left(\mathcal{K}_t,\mathcal{I}_t,{\mathcal{A}h}_t\right),\quad \forall t.
\end{equation}

The model of calendar aging is a nonlinear function: 
\begin{equation}\label{eq:cal_age}
\mathcal{U}_t^\mathrm{CAL}=G\left(\mathcal{S}_t\right)\ e^{\left(\frac{E_a}{R\ \mathcal{K}_t}\right)}\mathcal{Q}^{0.5}, \quad \forall t,
\end{equation}
\begin{align}\label{eq:cases_cal_age_G}
G\left(\mathcal{S}_t\right)=
    \begin{cases}
        a_1{\mathcal{S}_t}^2+a_2\mathcal{S}_t+a_3,    &0 \le\mathcal{S}_t \le 0.5Q\\
        b_1{\mathcal{S}_t}^2+b_2\mathcal{S}_t+b_3, &0.5Q < \mathcal{S}_t \le 0.7Q\\
        c_1{\mathcal{S}_t}^2+c_2\mathcal{S}_t+c_3, &0.7Q < \mathcal{S}_t \le 1.0Q
    \end{cases}
    ,\quad \forall t.
\end{align}
Equation (\ref{eq:cal_age}) is included in the problem by piece-wise linearization using binaries and auxiliary variables over the three SoC spans in (\ref{eq:cases_cal_age_G}).

The cycle aging model \cite{wang2014degradation} is also non-linear:
\begin{equation}
\mathcal{U}_t^\mathrm{CYC}=\left(q_1{\mathcal{K}_t}^2+q_2\mathcal{K}_t+q_3\right)e^{q_4\mathcal{I}_t}{Ah}_t, \quad \forall t.
\end{equation}
The cycle aging is linearized over the operation range of the battery, i.e., power: 0.1-1.0 MW, SoC: 10\%-90\%, and assuming a relatively constant temperature of 20$^\circ\mathrm{C}$ because stationary batteries are commonly kept in containers with regulated temperatures and effective battery management systems. 

Detail of mathematical linearization for calendar aging and cycle aging models is presented in \cite{10488452} and not repeated in this paper. The utilized aging parameters are presented in Table \ref{tab:bes_params}.

\section{Simulated Cases and Input Data}\label{sec:casestudy_data}

The impact of degradation is assessed for five market participation cases:
\begin{itemize}
    \item Case w/o FCR: do not participate in FCR,
    \item Case FCR-N: participate only in FCR-N,
    \item Case FCR-DU: participate only in FCR-D up,
    \item Case FCR-DD: participate only in FCR-D down,
    \item Case multi: multi-market participation is allowed at each hour.
\end{itemize}
The five cases are run twice, once when degradation cost is included in the objective function, and once without degradation cost. For the latter, the degradation cost is post-calculated and included in calculating the profit. The optimization problem is run for one day at a time and for a full year using a time-step of one minute.  

\subsection{Comparison metrics}
The results are assessed from two perspectives: monetary, and operation strategy. The monetary perspective compares the cases based on profit and degradation costs. The operation strategy compares the distribution of SOC, baseline power, and bid size. In addition, it compares the number of hours for participating in different combinations of markets.

\subsection{Input data}
The input data consists of the input parameters for the battery whose degradation model was formulated and the market-related data including frequency data and prices for the DA market, FCR-N, FCR-D up, FCR-D down, and imbalance. Battery input parameters are provided in Table \ref{tab:bes_params}, the battery data are taken from \cite{10488452}, prices are taken from \cite{ENTSOETransparencyPlatform,ESettOpenData}, and frequency data is taken from \cite{FingridOpenData}. Market zone SE3 and the year 2022 are used for prices and frequency data.
\begin{table}[ht]
\caption{Battery input parameters}
\label{tab:bes_params}
\begin{tabularx}{\linewidth}{@{}p{2.4cm}Xp{2.4cm}X@{}}
\toprule\toprule
Parameter                 & Value & Parameter                        & Value \\ \midrule
SoC range             & 10\%-90\% & Battery capacity             &   1 MWh    \\
Dis/Charge efficiency & 93\%      & Min, max dis/charge power &  0.0MW, 1.0MW     \\
Battery OM cost       &   $2\%\cdot137$ k€/MWh   & Battery replacement cost     &  137 k€/MWh     \\
EOL                   &   80\%     &   Interest rate                &  5\%    \\
Battery calendar lifetime     &  10 years  & Ratio of salvage cost &   0.5   \\
Temperature           &   20 $^{\circ}$ C  &&  \\ \midrule
\multicolumn{4}{>{\centering\hsize=\dimexpr6\hsize+6\tabcolsep+\arrayrulewidth\relax}X}{Cycle aging parameters\cite{wang2014degradation}}     \\ \midrule
\multicolumn{4}{>{\hsize=\dimexpr6\hsize+6\tabcolsep+\arrayrulewidth\relax}X}{At 20$^\circ\mathrm{C}$:\newline $q_1{\mathcal{K}_t}^2+q_2\mathcal{K}_t+q_3 = 0.0008$, $q4=0.3903$ }    \\ \midrule
\multicolumn{4}{>{\centering\hsize=\dimexpr6\hsize+6\tabcolsep+\arrayrulewidth\relax}X}{Calendar aging parameters\cite{wang2014degradation,calearo2019modeling}}\\ \midrule 
 \multicolumn{4}{>{\hsize=\dimexpr6\hsize+6\tabcolsep+\arrayrulewidth\relax}X}{$E_a = 24.5 \mathrm{kJ mol^{-1}}, R=8.314J \mathrm{mol^{-1}K^{-1}}$, $a_1=-1.1$, $a_2=89.7$, $a_3=1224.6$, $b_1=10.3$,$b_2=-1083.6$, $b_3=31447$, $c_1=2.6$, $c_2=-409.5$, $c_3=22035$} \\  
\bottomrule\bottomrule
\end{tabularx}
\end{table}

\section{Results and Discussion}\label{sec:results_discussion}
\begin{figure*}[htbp]
    \centering
    \includegraphics[width=\linewidth]{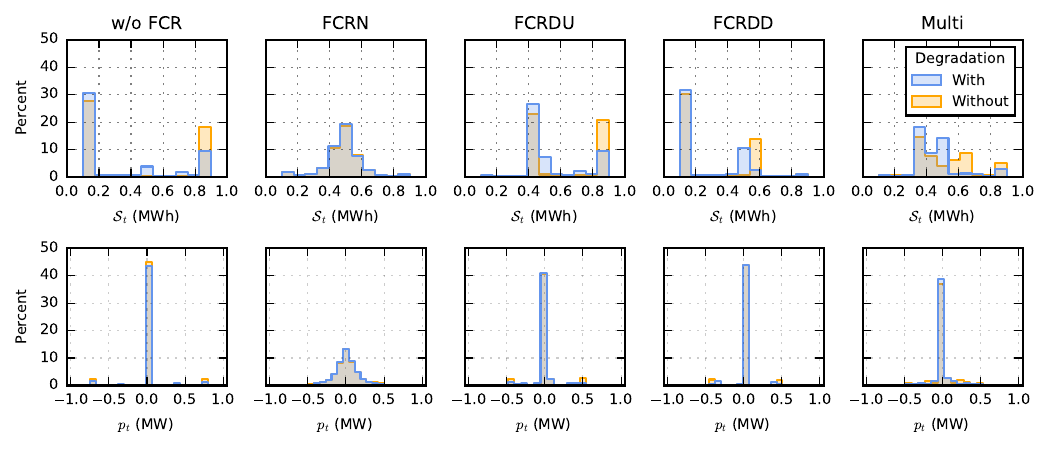}
    \caption{The battery utilization in different cases visualized by the histogram of battery SoE ($\mathcal{S}_t$) and power ($p_t$) as a percentage of all time step in a year}
    \label{fig:hist_soc_p}
\end{figure*}
Monetary results are presented in Table \ref{tab:monetary results}. Profit for cases without considering degradation in the objective function is obtained by post-calculating the aging cost and subtracting it from the objective function values. The results show that the multi-market case has a significantly larger profit compared to other market participation cases while having the second smallest degradation cost. Another observation is that considering degradation in the optimization does not have a considerable impact ($\mathrm{\le1\%}$) on the profit except in case w/o FCR. However, the total annual aging cost has been reduced by 5\%-29\% when battery degradation is considered in the optimization. This highlights that considering degradation in the objective function can provide operation strategies that not only result in a similar profit but also a longer lifetime for the battery. The utilization of the battery is visualized in Figure \ref{fig:hist_soc_p} for clarifying the reduction of battery degradation. The SoE and power data points show a shift towards smaller absolute values when degradation is considered. This explains the reduction of the aging costs when degradation is considered in the optimization.
\begin{table}[ht]
\caption{Annual monetary results in k€}
\label{tab:monetary results}
\begin{tabularx}{\linewidth}{@{}p{1.2cm}XXXX|XXXX|p{0.7cm}@{}}
\toprule\toprule
                       & \multicolumn{4}{c}{\textbf{W/o deg. in objective}} & \multicolumn{4}{c|}{\textbf{With deg. in objective}}& \multirow{2}={\textbf{$\Delta$ Cost tot. age.}} \\ \cmidrule(l){2-9} 
\multirow{-2}{*}{\textbf{Case}} & \textbf{Profit}   & \textbf{Cost cal. age. }  & \textbf{Cost cyc. age.} & \textbf{Cost tot. age.} & \textbf{Profit}  & \textbf{Cost cal. age.}  & \textbf{Cost cyc. age.}& \textbf{Cost tot. age.}& \\ \midrule
\textbf{w/o FCR} & 30 & 5.7 & 3.9 & 9.6 & 32 & 4.2 & 3.4 & 7.6 & -21\%\\
\textbf{FCR-N}   & 213  & 3.9 & 6.2 & 10.2 & 214  & 3.8 & 5.9 & 9.6 &  -5\% \\
\textbf{FCR-DU}  & 560  & 6.3 & 3.1 & 9.5 & 561  & 4.4 & 2.6 & 6.9 & -27\%\\
\textbf{FCR-DD}  & 303  & 3.5 & 2.7 & 6.2 & 303  & 2.6 & 2.3 & 4.9 & -21\%\\
\textbf{Multi}   & 706  & 4.5 & 3.0 & 7.5 & 708  & 3.1 & 2.2 &5.4 & -29\%\\ \bottomrule\bottomrule
\end{tabularx}
\end{table}

\begin{figure}[ht]
    \centering
    \includegraphics[width=\linewidth]{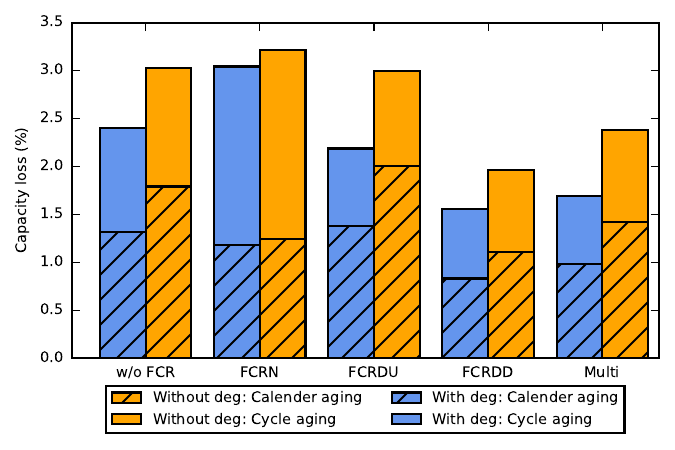}
    \caption{Distribution of annual battery capacity loss for the different market participation cases}
    \label{fig:capacity_loss}
\end{figure}

The total capacity loss of the battery over the year, $\sum_d\sum_t \mathcal{U}_{t,d}(\%)$, is presented in Figure \ref{fig:capacity_loss}. The largest degradation is expected in the FCR-N cases while the lowest degradation is expected for the FCR-D down cases. This is due to the large energy throughput and cycling for FCR-N cases while the FCR-D down cases require both lower cycling and SoC levels. In addition, FCR-N cases have the largest cycle aging costs while FCR-D up cases show the largest calendar aging. It is also worth noting that the multi-market participation cases have shown a lower degradation compared to the w/o FCR because the battery is utilized at lower SoC and power setpoints. 

\begin{figure*}[ht]
    \centering
    \includegraphics[width=1\linewidth]{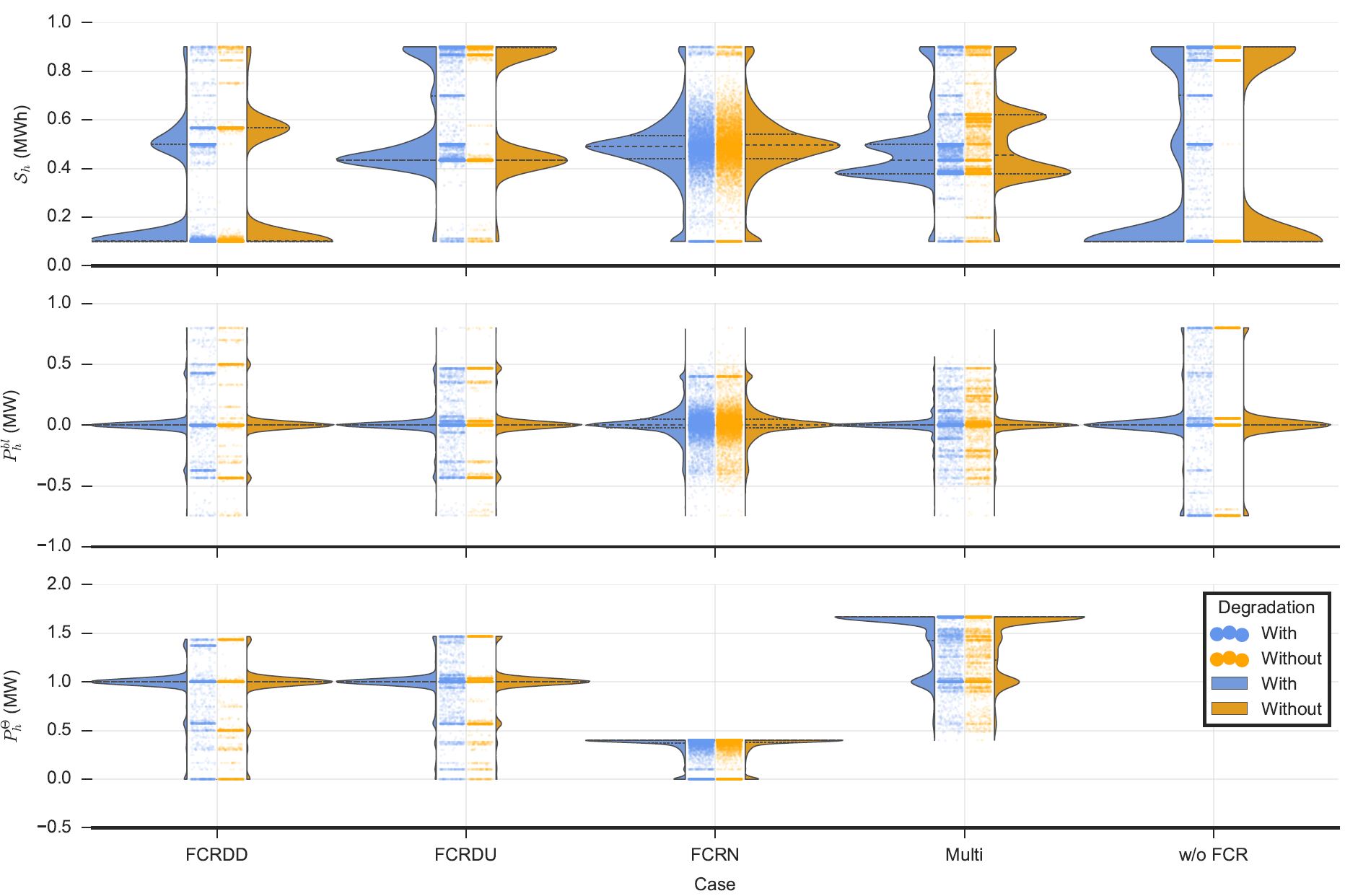}
    \caption{Operation strategy including the distribution of SoE at the beginning of each hour ($\mathcal{S}_h$), baseline power($p^{bl}_h$), and bid size ($p^\Theta_h$) for the simulated cases}
    \label{fig:strategy}
\end{figure*}

To better understand the difference in the operation strategies, the distribution of SoE in the beginning of each hour ($\mathcal{S}_h$), baseline power ($p^{bl}_h$), and bid size ($p^{\Theta}$) are presented in Figure \ref{fig:strategy}. The dashed lines show the first, second, and third quartiles. SoE at the beginning of each hour is considered as a part of the strategy because it is one of the main variables for satisfying the endurance requirements. For multi-market cases, the sum of bids in all the FCR markets is presented in Figure \ref{fig:strategy}. 

Several general observations can be made regarding the most dominant values for the strategy variables presented in Figure \ref{fig:strategy}. SoE at the beginning of each hour ($\mathcal{S}_h$) seems less discrete in the FCR-N case compared to the other cases. The dominant baseline power for all the cases is zero MW while the FCR-N case shows again a less discrete distribution. These two observations can highlight the potential complexities of real-life planning of an optimal FCR-N strategy compared to the other market participation cases which most likely boils down to the large energy throughput and a more regular activation of FCR-N services. 

The dominant bid sizes for cases FCR-D up and FCR-D down is 1 MW (Figure \ref{fig:strategy}). The larger bids were possible at hours with a non-zero reference power. The bids in the FCR-N case are limited to 0.4 MW due to considering technical requirements and the battery capacity. FCR-N is a symmetrical service with a required 1h endurance. Therefore, for example, if the SoE is at 0.5MWh, only 0.4MW can be provided for 1h in each direction considering the allowed SoC range of 10\%-90\%. For case Multi, the sum of the bids in all FCR markets are presented showing a dominant sum at 1.6 MW. This is the dominant optimal bidding strategy which is a simultaneous bid in FCR-D up and down markets (Table \ref{tab:bid_combination}). The sum is limited to 1.6 MW because of the technical requirement of power indicating a 20\% power availability in the opposite direction. The bid size results are yet another observation regarding the impact of considering technical requirements and their importance in obtaining realistic bidding and operation strategies.
\begin{table}[ht]
\caption{Choice of market as the number of hours in a year}
\label{tab:bid_combination}
\newcolumntype{Y}{>{\centering\arraybackslash}X}
\begin{tabularx}{\linewidth}{@{}lYYYYYYYY@{}}
\toprule\toprule
\multirow{2}{*}{\textbf{Case}} & \textbf{None} & \textbf{N} & \textbf{DU} & \textbf{DD} & \textbf{N+DU} & \textbf{N+DD}& \textbf{DU+DD}&\textbf{All} \\\cmidrule(l){2-9} 
                      & \multicolumn{8}{c}{\textbf{Without deg. in objective function}}  \\\cmidrule(l){2-9} 
\textbf{w/o FCR }              & 8760 & - & -    & -    & -    & -    & -     & -   \\
\textbf{FCR-N }                & 892  & 7868   & -     & -     & -     & -      & -      & -    \\
\textbf{FCR-DU}                & 73   & - & 8687 & -    & -    & -    & -     & -   \\
\textbf{FCR-DD}                & 279  & - & -    & 8481 & -    & -    & -     & -   \\
\textbf{Multi}                 & 0    & 3 & 1741 & 145  & 164  & 76   & 6110  & 521
\\ \cmidrule(l){2-9} 
                      & \multicolumn{8}{c}{\textbf{With deg. in objective function}}     \\\cmidrule(l){2-9}
\textbf{w/o FCR}               & 8760 & - & -    & -    & -    & -    & -     & -   \\
\textbf{FCR-N }                & 715  & 8045   & -     & -      & -      & -     & -      & -    \\
\textbf{FCR-DU}                & 70   & - & 8690 & -    & -    & -    & -     & -   \\
\textbf{FCR-DD}                & 202  & - & -    & 8558 & -    & -    & -     & -   \\
\textbf{Multi}                 & 0    & 2 & 1540 & 147  & 93   & 75   & 6382  & 521 
 \\ \bottomrule\bottomrule
\end{tabularx}
\end{table}

The mix of chosen markets in each case is presented in Table \ref{tab:bid_combination}. The inclusion of degradation in the objective function has led to a larger number of market participation hours. This is because the BESS can reduce calendar aging by regulating SoC. The change in the strategy comprises a larger number of participation hours but with a reduction in the number of hours that have the largest bidding power. This effect can be seen more clearly for the largest bids in cases FCR-D down and FCR-D up presented in Figure \ref{fig:strategy}. The dominant bid combination for case Multi is simultaneous FCR-DU and FCR-DD participation. This dominant bid combination besides the SoE levels in Figure \ref{fig:hist_soc_p} can clarify why the total battery degradation in Multi case is in between FCR-D up and FCR-D down cases (Figure \ref{fig:capacity_loss}).

It is worth noting that the model should not be seen as a ready-for-market bidding algorithm. The presented profit values are the theoretical maximum profit given the perspective of an oracle. In real-life, these profits might not be achieved to their full extent due to the lack of information and uncertainties in the input parameters. However, the model and its results can be interpreted as a benchmark indicating the maximum potential profit, and the potential impact of considering battery degradation in the optimization. Hence, the model can be used as an oracle model to evaluate and compare real-life bidding models that takes into account uncertainties in input parameters. 

\section{Conclusions}\label{sec:conclusions}
In this study, a novel MILP formulation was presented for Swedish multi-FCR participation including a detailed cycle and calendar aging besides the most recent technical requirements. The model was used for a full-year case study of the year 2022 to analyze the impact of degradation and technical requirements.

The results have shown that although considering degradation in the optimization does not have a significant impact on the profit, it can decrease the aging by 5\%-29\% leading to a more sustainable utilization of the battery. The results have also clearly demonstrated the impact and importance of fulfilling the technical requirements of the Swedish market. In addition, the highest profit could have been achieved in the multi-market participation that was dominated by simultaneous bidding in FCR-D up and down markets. The largest total degradation was for the dedicated FCR-N participation case while the lowest degradation was shown to be for the dedicated FCR-D down participation case.

The proposed model and the results can be utilized by flexibility asset owners to obtain more sustainable and yet profitable operation strategies. In addition, the formulation can be seen as an oracle model providing a maximum potential profit. This can be utilized as a benchmark for evaluating bidding models considering uncertainties.

\bibliographystyle{IEEEtran}

\begin{thebibliography}{10}
\providecommand{\url}[1]{#1}
\csname url@samestyle\endcsname
\providecommand{\newblock}{\relax}
\providecommand{\bibinfo}[2]{#2}
\providecommand{\BIBentrySTDinterwordspacing}{\spaceskip=0pt\relax}
\providecommand{\BIBentryALTinterwordstretchfactor}{4}
\providecommand{\BIBentryALTinterwordspacing}{\spaceskip=\fontdimen2\font plus
\BIBentryALTinterwordstretchfactor\fontdimen3\font minus \fontdimen4\font\relax}
\providecommand{\BIBforeignlanguage}[2]{{%
\expandafter\ifx\csname l@#1\endcsname\relax
\typeout{** WARNING: IEEEtran.bst: No hyphenation pattern has been}%
\typeout{** loaded for the language `#1'. Using the pattern for}%
\typeout{** the default language instead.}%
\else
\language=\csname l@#1\endcsname
\fi
#2}}
\providecommand{\BIBdecl}{\relax}
\BIBdecl

\bibitem{modig2022overview}
N.~Modig, R.~Eriksson, P.~Ruokolainen, J.~N. {\O}deg{\aa}rd, S.~Weizenegger, and T.~D. Fechtenburg, ``Overview of frequency control in the nordic power system,'' \emph{Nordic Analysis Group}, 2022.

\bibitem{8864014}
L.~Meng, J.~Zafar, S.~K. Khadem, A.~Collinson, K.~C. Murchie, F.~Coffele, and G.~M. Burt, ``Fast frequency response from energy storage systems—a review of grid standards, projects and technical issues,'' \emph{IEEE Transactions on Smart Grid}, vol.~11, no.~2, pp. 1566--1581, 2020.

\bibitem{hu2022potential}
Y.~Hu, M.~Armada, and M.~J. S{\'a}nchez, ``Potential utilization of battery energy storage systems (bess) in the major european electricity markets,'' \emph{Applied Energy}, vol. 322, p. 119512, 2022.

\bibitem{8999747}
T.~Zhao, A.~Parisio, and J.~V. Milanović, ``Distributed control of battery energy storage systems for improved frequency regulation,'' \emph{IEEE Transactions on Power Systems}, vol.~35, no.~5, pp. 3729--3738, 2020.

\bibitem{10045057}
Q.~Ma, W.~Wei, L.~Wu, and S.~Mei, ``Life-aware operation of battery energy storage in frequency regulation,'' \emph{IEEE Transactions on Sustainable Energy}, vol.~14, no.~3, pp. 1725--1736, 2023.

\bibitem{9445583}
E.~N. Sofia~Guzman, M.~Arriaga, C.~A. Cañizares, J.~W. Simpson-Porco, D.~Sohm, and K.~Bhattacharya, ``Regulation signal design and fast frequency control with energy storage systems,'' \emph{IEEE Transactions on Power Systems}, vol.~37, no.~1, pp. 224--236, 2022.

\bibitem{9102420}
A.~Oshnoei, M.~Kheradmandi, and S.~M. Muyeen, ``Robust control scheme for distributed battery energy storage systems in load frequency control,'' \emph{IEEE Transactions on Power Systems}, vol.~35, no.~6, pp. 4781--4791, 2020.

\bibitem{9103134}
P.~Astero and C.~Evens, ``Optimum operation of battery storage system in frequency containment reserves markets,'' \emph{IEEE Transactions on Smart Grid}, vol.~11, no.~6, pp. 4906--4915, 2020.

\bibitem{9832632}
I.~M. Casla, A.~Khodadadi, and L.~Söder, ``Optimal day ahead planning and bidding strategy of battery storage unit participating in nordic frequency markets,'' \emph{IEEE Access}, vol.~10, pp. 76\,870--76\,883, 2022.

\bibitem{9508854}
S.~Gao, H.~Li, J.~Jurasz, and R.~Dai, ``Optimal charging of electric vehicle aggregations participating in energy and ancillary service markets,'' \emph{IEEE Journal of Emerging and Selected Topics in Industrial Electronics}, vol.~3, no.~2, pp. 270--278, 2022.

\bibitem{9737408}
I.~Pavić, H.~Pandžić, and T.~Capuder, ``Electric vehicle aggregator as an automatic reserves provider under uncertain balancing energy procurement,'' \emph{IEEE Transactions on Power Systems}, vol.~38, no.~1, pp. 396--410, 2023.

\bibitem{subroto2022bess}
R.~K. Subroto, D.~Gebbran, A.~B. Moreno, and T.~Dragi{\v{c}}evi{\'c}, ``Bess optimal sizing and scheduling for energy arbitrage and frequency containment reserve via dual-loop optimization,'' in \emph{2022 IEEE Transportation Electrification Conference \& Expo (ITEC)}.\hskip 1em plus 0.5em minus 0.4em\relax IEEE, 2022, pp. 941--946.

\bibitem{thingvad2022economic}
A.~Thingvad, C.~Ziras, G.~Le~Ray, J.~Engelhardt, R.~R. Mosb{\ae}k, and M.~Marinelli, ``Economic value of multi-market bidding in nordic frequency markets,'' in \emph{2022 International Conference on Renewable Energies and Smart Technologies (REST)}, vol.~1.\hskip 1em plus 0.5em minus 0.4em\relax IEEE, 2022, pp. 1--5.

\bibitem{jacque2022influence}
K.~Jacqu{\'e}, L.~Koltermann, J.~Figgener, S.~Zurm{\"u}hlen, and D.~U. Sauer, ``The influence of frequency containment reserve on the cycles of a hybrid stationary large-scale storage system,'' \emph{Journal of Energy Storage}, vol.~52, p. 105040, 2022.

\bibitem{krupp2023operating}
A.~Krupp, R.~Beckmann, P.~Draheim, E.~Meschede, E.~Ferg, F.~Schuldt, and C.~Agert, ``Operating strategy optimization considering battery aging for a sector coupling system providing frequency containment reserve,'' \emph{Journal of Energy Storage}, vol.~68, p. 107787, 2023.

\bibitem{koltermann2023power}
L.~Koltermann, M.~C. Cort{\'e}s, J.~Figgener, S.~Zurm{\"u}hlen, and D.~U. Sauer, ``Power curves of megawatt-scale battery storage technologies for frequency regulation and energy trading,'' \emph{Applied Energy}, vol. 347, p. 121428, 2023.

\bibitem{entsoe_tech_req}
ENTSOe, ``Technical requirements for frequency containment reserve provision in the nordic synchronous area,'' ENTSOe, Tech. Rep., 2023.

\bibitem{InformationDifferentAncillary2024}
\BIBentryALTinterwordspacing
Information on different ancillary services. [Online]. Available: \url{https://www.svk.se/en/stakeholders-portal/electricity-market/provision-of-ancillary-services/information-on-different-ancillary-services/}
\BIBentrySTDinterwordspacing

\bibitem{10488452}
R.~Khezri, D.~Steen, E.~Wikner, and L.~A. Tuan, ``Optimal v2g scheduling of an ev with calendar and cycle aging of battery: An milp approach,'' \emph{IEEE Transactions on Transportation Electrification}, pp. 1--1, 2024.

\bibitem{wang2014degradation}
J.~Wang, J.~Purewal, P.~Liu, J.~Hicks-Garner, S.~Soukazian, E.~Sherman, A.~Sorenson, L.~Vu, H.~Tataria, and M.~W. Verbrugge, ``Degradation of lithium ion batteries employing graphite negatives and nickel--cobalt--manganese oxide+ spinel manganese oxide positives: Part 1, aging mechanisms and life estimation,'' \emph{Journal of Power Sources}, vol. 269, pp. 937--948, 2014.

\bibitem{ENTSOETransparencyPlatform}
\BIBentryALTinterwordspacing
{{ENTSO-E Transparency Platform}}. [Online]. Available: \url{https://transparency.entsoe.eu/}
\BIBentrySTDinterwordspacing

\bibitem{ESettOpenData}
\BIBentryALTinterwordspacing
{{eSett Open Data}}. [Online]. Available: \url{https://opendata.esett.com/}
\BIBentrySTDinterwordspacing

\bibitem{FingridOpenData}
\BIBentryALTinterwordspacing
Fingrid {{Open Data}}. [Online]. Available: \url{https://data.fingrid.fi/en}
\BIBentrySTDinterwordspacing

\bibitem{calearo2019modeling}
L.~Calearo, A.~Thingvad, and M.~Marinelli, ``Modeling of battery electric vehicles for degradation studies,'' in \emph{2019 54th International Universities Power Engineering Conference (UPEC)}.\hskip 1em plus 0.5em minus 0.4em\relax IEEE, 2019, pp. 1--6.

\end{thebibliography}

\end{document}